\documentclass[twoside]{article}
\usepackage{fleqn,espcrc2}

\usepackage{graphicx}

\newcommand{\mstr}{\mbox{$M_{\rm str}$}}                              
\newcommand{\nue}{\mbox{$\nu_e$}} 
\newcommand{\num}{\mbox{$\nu_\mu$}}                                            
\newcommand{\nut}{\mbox{$\nu_\tau$}}                                            
\newcommand{\nus}{\mbox{$\nu_s$}}                                            
\newcommand{\stt}{\mbox{$\sin^2 2 \theta$}}   
\newcommand{\dms}{\mbox{$\Delta m^2$}}

\newcommand{\skipblk}[1]{}                                                      
\newcommand{\beqa}{\begin{eqnarray}}
\newcommand{\eeqa}{\end{eqnarray}} 
\newcommand{\bqa}{\begin{eqnarray}}
\newcommand{\eqa}{\end{eqnarray}}  

\newcommand{\doub}[3]{\mbox{$                                                   
\left( \begin{array}{c} #1    \\ #2  \end{array} \right)_#3$}}                  

\newcommand{\PR}[3]{{Phys. Rev.} {\bf #1}, #2 (19#3)}                             
\newcommand{\PL}[3]{{Phys. Lett.} {\bf #1}, #2 (19#3)}                            
\newcommand{\NP}[3]{{Nucl. Phys.} {\bf #1}, #2 (19#3)}                            
\newcommand{\PRL}[3]{{Phys. Rev. Lett.} {\bf #1}, #2 (19#3)}

\newcommand{\ie}{{\em i.e., }}                                                  
\newcommand{\x}{\mbox{$\times$}}                                                

\newcommand{\beq}{\begin{equation}}                                             
\newcommand{\eeq}{\end{equation}}                                               

\newcommand{\RA}{\mbox{$\rightarrow$}}

\def\mxth{\mathsurround=0pt }
\def\xversim#1#2{\lower2.pt\vbox{\baselineskip0pt \lineskip-.5pt
  \ialign{$\mxth#1\hfil##\hfil$\crcr#2\crcr\sim\crcr}}}             
\def\simgr{\mathrel{\mathpalette\xversim >}}                                    
\def\simle{\mathrel{\mathpalette\xversim <}}

\title{Neutrino Oscillation Workshop 2000: Conference Summary}

\author{Paul Langacker\address{Department of Physics and Astronomy \\ 
          University of Pennsylvania, Philadelphia PA 19104-6396, USA}}
       
\begin{document}

\begin{abstract}The NOW2000 Workshop summarized the present status and future 
possibilities for all aspects of neutrino physics and astrophysics.  Neutrino
oscillation physics has truly come of age in the last few years. It is
now data driven (analogous to cosmology and much of high energy physics in general).
Experimental techniques and the theoretical interpretation have developed
dramatically. An example of the latter includes the more realistic
analysis of neutrino oscillations in the framework of 3 or more mass eigenstates. The
phenomenological emphasis has shifted from degenerate to hierarchical neutrino spectra. 
\vspace{1pc}
\end{abstract}

\maketitle

\section{OUTLINE}
\begin{itemize}
\item Neutrinos as a Probe
\item Theoretical Framework
\item Cosmology (BBN, CMB, $\Delta B$, $\Delta L$, relic)
\item Violent Astrophysical Events (GRBs, AGNs, supernovae)
\item Low Energy Neutrinos (Astrophysical and Laboratory)
\item Outlook
\end{itemize}

\section{NEUTRINOS AS A PROBE}
Neutrinos are a unique probe of many aspects of physics and
astrophysics, on scales ranging from $10^{-33}$ to $10^{+28}$ cm.
 Particle physics applications include:
  \begin{itemize}
   \item $\nu N, \mu N, e N$ scattering: existence/ properties
of quarks, QCD
  \item Weak decays ($n \rightarrow p e^- \bar{\nu}_e, \mu^- 
\rightarrow e^- \nu_\mu \bar{\nu}_e$): Fermi theory, parity
violation,  mixing
  \item Neutral current, $Z$-pole, atomic parity: electroweak
unification, field theory, $m_t$; severe constraint on physics to
TeV scale
\item Neutrino mass: constraint on TeV physics, grand
unification, superstrings, extra dimensions
  \end{itemize}
Similarly, their relevance to astrophysics and cosmology includes
  \begin{itemize}
   \item Core of Sun
   \item Supernova dynamics
   \item Atmospheric neutrinos (cosmic rays)
   \item Violent events: GRBs, AGNs, cosmic rays
   \item Large scale structure (dark matter)
   \item Nucleosynthesis (big bang: small $A$; stellar: to
iron; supernova: large $A$)
   \item Baryogenesis
  \end{itemize}
Astrophysical applications are especially challenging, because one must
often simultaneously disentagle the properties of the neutrinos and
of the astrophysical sources.

\section{THEORETICAL FRAMEWORK}

There are a confusing variety of models of neutrino mass. Here,
I give a brief survey of the principle classes and of some of the terminology. 
For more
detail,  see~\cite{lr0,feruglio}.

\subsection{Weyl, Dirac, and Majorana neutrinos}
\label{theorysection}

A Weyl two-component spinor is a left ($L$)-handed\footnote{The subscripts
$L$ and $R$ really refer to the left and right chiral projections. In
the limit of zero mass these correspond to left and right helicity
states.} \ particle state, $\psi_L$, which is necessarily associated by CPT
with a right ($R$)-handed antiparticle state\footnote{Which is referred
to as the particle or the antiparticle is a matter of convenience.}
\ $\psi^c_R$. One refers to active (or ordinary) neutrinos
as left-handed neutrinos
which transform as $SU(2)$ doublets with a charged lepton
partner. They therefore have normal weak interactions, as do their right-handed
anti-lepton partners, 
\beq \doub{\nu_e}{e^-}{L} \stackrel{\rm  CPT}{\longleftrightarrow}
\doub{e^+}{\nu^c_e}{R}. \eeq
Sterile\footnote{Sterile neutrinos are often referred to as
``right-handed'' neutrinos, but that terminology is
confusing and inappropriate when Majorana masses are present.}
\ neutrinos are  $SU(2)$-singlet neutrinos, which can be added to the 
standard model and  are predicted in most extensions. They have
no ordinary weak interactions except those induced by mixing with active
neutrinos. It is usually convenient to define the $R$ state as the
particle and the related $L$ anti-state as the antiparticle.
\beq N_R \stackrel{\rm  CPT}{\longleftrightarrow} N^c_L. \eeq
(Sterile neutrinos will sometimes also be denoted $\nu_s$.)

Mass terms describe transitions between right ($R$)
and left ($L$)-handed states.
A Dirac mass term, which conserves lepton number, involves transitions
between two distinct Weyl neutrinos
$\nu_L$ and $N_R$:
\beq
- L_{\rm Dirac}  =  m_D (\bar{\nu}_L N_R +
\bar{N}_R \nu_L)
  =  m_D \bar{\nu} \nu, \eeq
where the Dirac field is defined as $\nu \equiv \nu_L + N_R$.  Thus a
Dirac neutrino has four components $ \nu_L, \; \nu_R^c, \; N_R, \;
N_L^c$,
and the mass term allows a conserved lepton number $L = L_\nu
+ L_N$.  This and other types of mass terms can easily be generalized
to three or more families, in which case the masses become matrices.
The charged current transitions then involve
a leptonic mixing matrix (analogous to the
Cabibbo-Kobayashi-Maskawa (CKM) quark mixing  matrix), which
can lead to neutrino oscillations between the light neutrinos.

For an ordinary Dirac neutrino
the $\nu_L$ is active 
and the $N_R$ is
sterile.
The transition is $\Delta I = \frac{1}{2}$,
where $I$ is the weak isospin.  The mass requires $SU(2)$ breaking and
is generated by a Yukawa coupling
\beq  -L_{\rm Yukawa} = h_\nu (\bar{\nu}_e \bar{e})_L
\left( \begin{array}{c} \varphi^0 \\ \varphi^- \end{array} \right)
  N_R + H.C. \eeq
One has $m_D = h_\nu v/\sqrt{2}$,
where the vacuum expectation value (VEV) of
the Higgs doublet is
$ v = \sqrt{2} \langle \varphi^o \rangle = ( \sqrt{2} G_F)^{-1/2} =
246$ GeV,
and $h_\nu$ is the Yukawa coupling.
A Dirac mass is just like the quark and charged lepton masses, but
that leads to the question of    why it is so small: one  requires
$h_{\nu_e} < 10^{-11}$  to have $m_{\nu_e} < 1$ eV.  

A Majorana mass, which violates lepton number by two units $(\Delta L
= \pm 2)$, makes use of the right-handed antineutrino, 
$\nu^c_R$, rather than a separate Weyl neutrino.  It is a transition
from an antineutrino into a neutrino. Equivalently, it can be viewed
as the creation  or annihilation of two neutrinos, and if present
it can therefore lead to neutrinoless double beta decay.
The form of a Majorana mass term is
\beqa - L_{\rm Majorana}  & = &\frac{1}{2} m_T (\bar{\nu}_L \nu_R^c +
\bar{\nu}^c_R \nu_L )   = \frac{1}{2} m_T \bar{\nu} \nu \nonumber \\
& = & \frac{1}{2} m_T (\bar{\nu}_L C \bar{\nu}_L^T + H.C.),
 \eeqa
where $\nu = \nu_L +\nu^c_R$ is a self-conjugate two-component state
satisfying $\nu = \nu^c = C \bar{\nu}^T$, where $C$ is the
charge conjugation matrix.  If $\nu_L$ is active then
$\Delta I = 1$ and $m_T$ must be generated by either an elementary Higgs
triplet or by an effective operator involving two Higgs doublets
arranged to transform as a triplet. 

One can also have a Majorana
mass term
\beq - L_{\rm Majorana} = \frac{1}{2} m_N (\bar{N}^c_L N_R +
\bar{N}_R N^c_L ) \eeq
for a sterile neutrino. This has $\Delta I = 0$ and thus
can be generated by the VEV of a
Higgs  singlet\footnote{In principle this could also be generated by a bare
mass, but this is usually forbidden by higher symmetries in extensions 
of the standard model.}.

\subsection{Models of neutrino mass} \label{modelsmass}

Almost all extensions of the standard model lead to nonzero neutrino
masses at some level, often in the observable ($10^{-5}-10$ eV) range.
One should therefore view neutrino mass as {\it top-down} physics. e.g., one
can hope to compare the predictions of
a specific superstring, GUT, or other model with the observed spectra, but
it is hard to work backwards and infer the underlying theory from the observations.
There are large numbers of models of neutrino mass. 
Some of the principle classes and general issues are:
\begin{itemize}
\item A triplet majorana mass $m_T$ can be generated by
the VEV $v_T$ of a Higgs triplet field. Then, $m_T = 
h_T v_T$, where $h_T$ is the relevant Yukawa
coupling. Small values of $m_T$ could be due to a small scale
$v_T$, although that introduces a new hierarchy problem.
The simplest implementation
is the Gelmini-Roncadelli (GR) model~\cite{lr19},
in which lepton number is spontaneously broken by $v_T$. The
original GR model is now excluded by the LEP data on the $Z$
width.

\item
A very different class of models are those in which the neutrino
masses are zero at the tree level (typically because no sterile neutrino
or elementary Higgs triplets are introduced), but only generated by
loops \cite{lr57}, \ie
radiative generation.  Such models
generally require
the {\em ad hoc} introduction of new scalar particles at the TeV scale
with nonstandard
electroweak quantum numbers and lepton number-violating couplings.
They have also been introduced in an attempt to generate large electric or
magnetic dipole moments. 

\item In the seesaw models~\cite{lr16}, a small Majorana mass
is induced by mixing between an active neutrino and a very heavy
Majorana sterile neutrino $M_N$. The light (essentially active)
state has a naturally small mass
\beq m_\nu \sim \frac{m_D^2}{M_N} \ll m_D. \eeq
There are literally hundreds of seesaw models, which differ in the scale
$M_N$ for the heavy neutrino (ranging from the TeV scale to grand unification
scale), the Dirac mass $m_D$ which connects the ordinary and sterile
states and induces the mixing (e.g., $m_D \sim m_u$ in most grand unified theory
(GUT) models, or $\sim m_e$ in left-right symmetric models), the patterns
of $m_D$ and $M_N$ in three family generalizations, etc. 

\item There are many mechanisms for neutrino mass generation in
 supersymmetric models with $R$ parity
breaking~\cite{susynu}. Breaking induced  by bilinear terms connecting Higgs and lepton
doublets in the superpotential or by the expectation values of scalar neutrinos
leads to seesaw-type mixings of neutrinos with
 heavy neutralinos. 
Cubic $R$
parity violating terms lead to loop-induced neutrino masses.

\item Grand unified theories  are excellent candidates for seesaw models,
with $M_N$ at or a few orders of magnitude below the unification scale.
In addition to the gauge and Yukawa unification, realistic models for the quark and
charged-lepton masses generally involve complicated Higgs structures and additional
family symmetries to constrain the Yukawa couplings (fermion textures),
often leading to interesting predictions for the neutrino spectrum. 
Detailed $SO(10)$ models were described by Raby~\cite{raby}. The most predictive
versions are excluded, while generalizations have less predictive power.
It is
difficult to embed such structures into superstring models.

\item Heterotic superstring models often predict the existence of higher-dimensional
(nonrenormalizable) operators (NRO) such as
\beq -L_{\rm eff} = \bar{\psi}_L H \left(\frac{S}{\mstr}\right)^P \psi_R
+ H.C., \eeq
where $H$ is the ordinary Higgs doublet, $S$ is a new scalar field which is
a singlet under the standard model gauge group, and $\mstr \sim
10^{18}$ GeV is the string scale. In many cases $S$ will acquire
an intermediate scale VEV (e.g., $10^{12}$ GeV), leading to an
effective Yukawa coupling 
\beq h_{\rm eff} \sim v \left(\frac{\langle S \rangle}{\mstr}\right)^P
\ll v. \eeq
Depending on the dimensions $P$ of the various operators and on
the scale $\langle S \rangle$, it may be possible to generate
an interesting hierarchy for the quark and charged lepton masses
and to obtain naturally small Dirac neutrino masses~\cite{nro}.
Similarly, one may obtain triplet and singlet Majorana neutrino
masses, $m_T$ and $m_N$ by analogous higher-dimensional operators.
The former are generally too small to be relevant. 
Depending on the operators
the latter ($m_N$) may be absent, implying small Dirac masses; 
small, leading to the possibility of significant mixing between ordinary
and sterile neutrinos~\cite{sterile}; or
large, allowing a conventional seesaw. 

\item There are many models in which large extra dimensions affect
the neutrino masses, generally involving singlet neutrinos
propagating in the bulk and/or coupling with Kaluza-Klein excitations.
These couplings could even lead to oscillations without neutrino
masses~\cite{dienes,lusignoli}. 

\item Realistic models for three or more neutrinos often involve fermion
textures~\cite{feruglio,joaquim,tanimoto}, i.e., particular forms for the fermion mass
matrices involving zeroes or hierarchies of non-zero entries. Such textures are  usually
assumed to be  due to broken family symmetries. However, they could also be associated
with unknown dynamics, such as string selection rules in an underlying theory.

\item It is often assumed that small neutrino masses are most likely Majorana.
This is expected in the simplest seesaw models. However, the possibility of
small Dirac masses should not be excluded. These can easily come about
in (string-motivated) intermediate scale models, and possibly in
loop-induced scenarios.

\item Mixed models, in which comparable Majorana
and Dirac mass terms are both present, will be further discussed
in the next section.

\end{itemize}

\subsection{Light sterile neutrinos}

Most extensions of the standard model predict the existence of sterile
neutrinos. For example, simple $SO(10)$ and $E_6$ grand unified theories
predict one or two sterile neutrinos per family, respectively. The only real
questions are whether the ordinary and sterile neutrinos of the same chirality mix
significantly with each other,   whether some or all of 
the mass eigenstate neutrinos are
sufficiently light, and how many
 (e.g., whether there are 1, 3, 6 or some other number of
light sterile neturinos).

 When there are only Dirac masses, the
ordinary and sterile states do not mix because of the conserved lepton number. Pure
Majorana masses do not mix the ordinary and sterile sectors either.
In the seesaw model the mixing is negligibly small, and the 
(mainly) sterile eigenstates are too heavy to be relevant to oscillations.
The only way to have significant mixing and  small mass eigenstates
is for the Dirac and Majorana neutrino mass terms to be extremely
small and to also be comparable to each
other.
This appears to
require two miracles in conventional models of neutrino mass.

One promising possibility involves the generation of neutrino
masses from higher-dimensional operators in theories involving
an intermediate scale~\cite{nro},
as described in Section \ref{modelsmass}.
Another~\cite{parallel} involves sterile neutrinos
associated with a parallel hidden sector of nature, as suggested in
some superstring and supergravity theories. Yet another associates
the sterile neutrino with the light ($10^{-3}$ eV) Kaluza-Klein modes associated
with a large (mm) extra dimension~\cite{cmy}.

\begin{figure}[h]
\includegraphics*[scale=0.58]{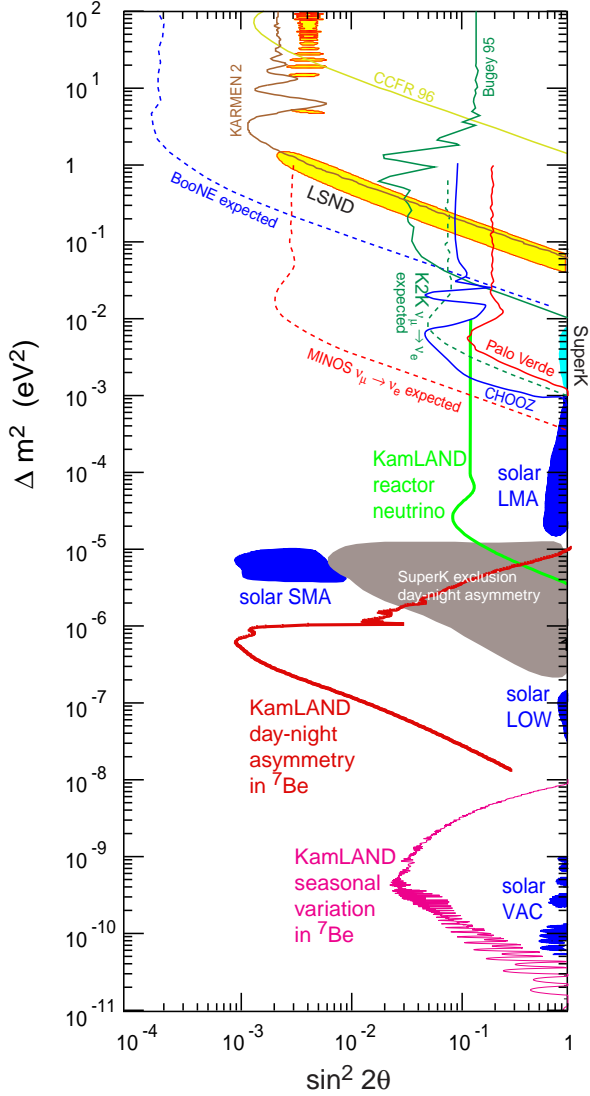}
\caption{Neutrino oscillation plot, indicating the solar neutrino small
mixing angle (SMA), large mixing angle (LMA), low mass (LOW), and vacuum (VAC)
solutions, as well as the Superkamiokande atmospheric neutrino range, the LSND
region, and various exclusion regions. From~\cite{albright}.}
\label{bigpicture}
\end{figure}

\subsection{Mass and mixing patterns}
\label{global}
Various scenarios for the neutrino spectrum are possible, depending
on which of the experimental indications one accepts~\cite{akhmedov}. 

\subsubsection{3 neutrino schemes~\cite{lisi,kajita,gavela,minakata,lindner,donini}}
The simplest
scheme, which accounts for the Solar (S) and Atmospheric (A) neutrino
results, is that there are just three light neutrinos, all active,
and that the mass eigenstates $\nu_i$ have masses in a hierarchy,
analogous to the quarks and charged leptons. In that case, the
atmospheric and solar neutrino mass-squared differences are measures
of the mass-squares of the two heavier states, so that
$m_3 \sim (\Delta m^2_{\rm atm})^{1/2} \sim 0.03-0.1$ eV;
$m_2 \sim (\Delta m^2_{\rm solar})^{1/2} \sim 0.003$ eV (for MSW)
or $\sim 10^{-5}$ eV (vacuum oscillations), and $m_1 \ll m_2$.
The weak eigenstate neutrinos
$\nu_a =(\nu_e, \nu_\mu, \nu_\tau)$ are related to the mass eigenstates 
$\nu_i$ by a unitary transformation $\nu_a = U_{ai} \nu_i$.
If one makes the simplest assumption (from the Superkamiokande, CHOOZ, and Palo Verde
 data),
that the $\nu_e$ decouples entirely from the atmospheric neutrino oscillations,
$U_{e3}=0$,
(of course, one can relax this assumption somewhat) and ignores possible 
CP-violating phases, then 
\beqa
\left(  
\begin{array}{c}
\nu_e \\ \nu_\mu \\ \nu_\tau
\end{array}
\right)
& = &
\left(  
\begin{array}{ccc}
1 & 0 & 0 \\
0 & c_\alpha & -s_\alpha  \\
0 & s_\alpha  & c_\alpha
\end{array}
\right)  \nonumber \\
 & \x &
\left(  
\begin{array}{ccc}
c_\theta & -s_\theta & 0 \\ 
s_\theta & c_\theta & 0 \\
0 & 0 & 1
\end{array}
\right)
\left(  
\begin{array}{c}
\nu_1 \\ \nu_2 \\ \nu_3
\end{array}
\right),
\eeqa
where $\alpha$ and $\theta$ are mixing angles associated with the atmospheric
and solar neutrino oscillations, respectively, and where $c_\alpha \equiv 
\cos \alpha$, $s_\alpha \equiv \sin \alpha$, and similarly for $c_\theta, s_\theta$.

For maximal atmospheric neutrino mixing, $\sin^2 2 \alpha \sim 1$, this implies
$c_\alpha = s_\alpha = 1/\sqrt{2}$, so that
\beq
U = 
\left(  
\begin{array}{ccc}
c_\theta & -s_\theta & 0 \\ 
\frac{s_\theta}{\sqrt{2}} & \frac{c_\theta}{\sqrt{2}} & -\frac{1}{\sqrt{2}} \\
\frac{s_\theta}{\sqrt{2}} & \frac{c_\theta}{\sqrt{2}} & \frac{1}{\sqrt{2}}
\end{array}
\right).
\eeq
For small $\theta$, this implies that $\nu_{3,2} \sim \nu_{+,-} \equiv (\nu_\tau \pm
\nu_\mu)/\sqrt{2}$ participate in atmospheric oscillations, 
while the solar neutrinos
are associated with a small additional mixing between $\nu_e$ and $\nu_-$.
Another limit, suggested by the possibility of vacuum oscillations for
the solar neutrinos, is $\sin^2 2 \theta \sim 1$, or
$c_\theta = s_\theta = 1/\sqrt{2}$, yielding
\beq
U = 
\left(  
\begin{array}{ccc}
\frac{1}{\sqrt{2}} & -\frac{1}{\sqrt{2}} & 0 \\ 
\frac{1}{2} & \frac{1}{2} & -\frac{1}{\sqrt{2}} \\
\frac{1}{2} & \frac{1}{2} & \frac{1}{\sqrt{2}}
\end{array}
\right),
\eeq
which is referred to as  bi-maximal mixing~\cite{smirnov,palazzo,jezabek}.
A number of authors have discussed this pattern and how it might
be obtained from models, as well as how
much freedom there is to relax the assumptions of maximal
atmospheric and solar mixing (the data actually allow $\sin^2 2 \alpha \simgr
0.8$ and $\sin^2 2 \theta \simgr 0.6$) or the complete decoupling of $\nu_e$
from the atmospheric neutrinos. Another popular pattern,
\beq
U = 
\left(  
\begin{array}{ccc}
\frac{1}{\sqrt{2}} & -\frac{1}{\sqrt{2}} & 0 \\ 
\frac{1}{\sqrt{6}} & \frac{1}{\sqrt{6}} & -\frac{2}{\sqrt{6}} \\
\frac{1}{\sqrt{3}} & \frac{1}{\sqrt{3}} & \frac{1}{\sqrt{3}}
\end{array}
\right),
\eeq
known  as democratic mixing~\cite{jezabek}, yields maximal solar oscillations
and near-maximal ($8/9$) atmospheric oscillations.

The atmospheric neutrino data only determine the magnitude of $|\Delta m^2_{32}|
\equiv |m_3^2 -
m_2^2| = \Delta m^2_{\rm atm}$, so there is a variant inverted hierarchy in which
$m_3 < m_1 < m_2$, with $m_2^2 - m_1^2
= \Delta m^2_{\rm solar} \ll m_2^2$, i.e., there is a quasi-degeneracy\footnote{For
MSW oscillations, the data implies that most likely
$m_2^2 > m_1^2$ under the convention
that $\theta < \pi/4$. For vacuum oscillations, the sign of
$\Delta m^2_{\rm solar}$ is undetermined.},
and $m_2^2 - m_3^2
= \Delta m^2_{\rm atm}$.

In the hierarchical and inverted patterns, the masses are all too small
to be relevant to mixed dark matter, in which one of the components of
the dark matter is hot, i.e., massive neutrinos.
However,
the solar and atmospheric oscillations only determine the differences in
mass squares, so there are variants on these scenarios in which the three
mass eigenstates are nearly degenerate rather than
hierarchical,
with small splittings associated with $\Delta m^2_{\rm atm}$
and $\Delta m^2_{\rm solar}$. For the common mass $m_{\rm av}$
in the 1 -- several eV range, the hot dark matter could be relevant
 on large scales. Such mixed dark matter (with another, larger, component of cold
dark matter accounting for smaller structures) models~\cite{masiero}
 were once quite popular. 
However, recent evidence for dark energy (cosmological constant or quintessence)
eliminates most of the motivation for considering degenerate models, 
although they are still
not excluded. Another problem with both the degenerate and inverted schemes is that
the near degeneracies are usually unstable with respect to radiative
corrections~\cite{lola}.

If the neutrinos
are Majorana they could also lead to neutrinoless double beta decay, $\beta
\beta_{0\nu}$~\cite{klapdor}. Current limits imply an upper limit of
\beq \langle m_{\nu_e} \rangle = \sum_i \eta_i U_{ei}^2 |m_i|
\simle 0.35\ {\rm eV}, \label{mnueff} \eeq
on the effective mass for a mixture of light Majorana mass eigenstates,
where $\eta_i$ is the CP-parity (or phase) of $\nu_i$. There is an additional
uncertainty on the right  due to the nuclear matrix elements.
(There is no constraint on Dirac neutrinos.)  This constraint
is very important for the degenerate scheme for $m_{\rm av}$
in the eV range.
The combination of small $\langle m_{\nu_e} \rangle \ll m_{\rm av}$,
maximal
atmospheric mixing, and $U_{e3}=0$ would imply cancellations, 
so that $\eta_1 \eta_2 =
-1$ and $c_\theta = s_\theta = 1/\sqrt{2}$, i.e., maximal solar mixing.
However, there is  room to relax all of these assumptions considerably.
For the hierarchical and inverted schemes 
$\langle m_{\nu_e} \rangle$ is small compared to the current experimental limit.
However, proposed future experiments could be sensitive to the inverted 
case. 

\subsubsection{4 neutrino schemes~\cite{lisi,giunti,peres,marrone}}
\label{scenarios}

The LSND results~\cite{spentzouris}, if confirmed, would almost certainly imply a fourth,
sterile, neutrino $\nu_s$ (the $Z$ lineshape does not allow a fourth light active neutrino),
in which one or two of the neutrinos are separated from the others by
$\sqrt{\Delta m^2_{\rm LSND}} \sim 0.4 - 1$ eV.
There could be even more sterile neutrinos.

In the $3 + 1$ schemes, $\nu_4$ is heavier or lighter than $\nu_{1,2,3}$
by $\sqrt{\Delta m^2_{\rm LSND}}$, with the splittings between the 
latter controlled by $\Delta m^2_{\rm atm}$ and $\Delta m^2_{\rm solar}$.
This  case has generally been considered excluded by limits from
$\nu_e$ and $\nu_\mu$ disappearance. However, small recent changes in the LSND 
favored range (to lower \dms)
imply that these schemes are barely allowed for
$\nu_4 \sim \nu_s$, and possibly for $\nu_4 \sim \nu_\tau$.
The $\nu_4 \sim \nu_s$ case may offer a theoretical advantage over $2 + 2$ schemes
in that the $\nu_s$ is more distinct from the active neutrinos.

In the $2+2$ schemes, 
one has two pairs of mass eigenstates $\nu_{1,2}$
and
$\nu_{3,4}$, with
$\pm \Delta m^2_{34} \sim  \Delta m^2_{\rm atm} \sim 10^{-3}-10^{-2}$ eV$^2$, 
$\Delta m^2_{12} \sim \Delta m^2_{\rm solar} \sim 10^{-5}$ eV$^2$ (MSW) or $10^{-10}$
eV$^2$ (vacuum), and
$\pm \Delta m^2_{24} \sim  \Delta m^2_{\rm LSND} \sim 0.2-1$ eV$^2$,
where $\Delta m^2_{ij} \equiv m_j^2 - m_i^2$.
The reactor data imply that $\nu_e$ must be largely restricted to one of the
pairs. The cases $\Delta m^2_{24} > 0$ and $<0$ are referred to
as hierarchical and inverted, respectively. The inverted case, and to a
somewhat lesser extent the hierarchical case, are quasi-degenerate, and
may be unstable under radiative corrections~\cite{lola}.

The $2+2$ and some versions of the $3+1$ models involve a 
significant hot or warm neutrino component to the dark matter.
The extra sterile neutrino may be of importance for big bang nucleosynthesis.
The versions with $\nu_1$ in the heavier group may give significant
contributions to $\beta \beta_{0\nu}$, although there may be major cancellations
for large mixing.

The recent SuperKamiokande~\cite{kajita} and MACRO~\cite{ronga} atmospheric neutrino data
exclude the pure $\nu_{\mu,s} \sim \nu_{3,4}$ case, in which the atmospheric neutrino results
are associated with $\nu_\mu \RA \nu_s$, while Super-K solar neutrino data~\cite{suzuki}
 probably eliminates
the pure $\nu_e \RA \nu_s$ (i.e., $\nu_{e,s} \sim
 \nu_{1,2}$) explanation for the solar neutrinos. These were the simplest and perhaps most
plausible cases. However, more general mixing schemes with significant $\nu_s$
admixtures in the solar and atmospheric neutrinos are 
possible~\cite{lisi,giunti,garcia}.

\subsection{Alternatives to neutrino oscillations}
Alternative explanations of the atmospheric and solar neutrino anomalies were
reviewed by Lusignoli~\cite{lusignoli}.

For the atmospheric neutrinos, flavor changing neutral currents (FCNC)~\cite{guzzo}, 
Lorentz
invariance violation (LIV), and Equivalence Principle Violation 
(EPV)~\cite{kuo,nunokawa} models
for neutrino mixing have been suggested as alternatives to conventional
flavor mixing. However, they are not consistent with zenith
or upward throughgoing distributions. For example, LIV and EPV
predict an $LE^n$ dependence with $n \ne -1$, contrary to the
SuperK data. $L$ and $E$ refer to the distance travelled and
energy of the $\nu_\mu$, respectively.  Pure $\nu_\mu$ decay
models are not viable, but a hybrid oscillation and decay model with two
nearly degenerate states, the lightest decaying to sterile states, cannot be
excluded. Also, decoherence models, in which the coherence between two
states is lost by some unknown mechanism (e.g., interaction with
quantum foam), are consistent with the
data.

There are viable descriptions of the solar neutrino data involving involving
FCNC, LIV, and EPV. Another alternative  is resonant
spin flavor precession (RSPF)~\cite{lusignoli,akhmedov,broggini}, 
in which $\nu_{eL}$ is
transformed into a sterile or active right-handed state. This requires
a much larger neutrino magnetic moment $\mu_\nu$ (which can be Dirac for a sterile
$N_R$ final state, or a Majorana transition moment for an active $\nu^c_{\mu R}$ or
$\nu^c_{\tau R}$) than is predicted by most models and
a rather large solar magnetic field. The original motivation for RSPF
for magnetic transitions was an apparent time dependence in the Homestake
data. This is no longer wanted experimentally, but time independent
magnetic effects could still be relevant for transitions occurring below the
convective zone. $\nu_{eL} \RA \nu^c_{eR}$ can be generated by
combined magnetic and flavor-oscillation effects.

\subsection{What is needed?}
For the future, one wants to establish neutrino oscillations (or constrain small
admixtures of other effects in hybrid scenarios), establish which mixing scenarios
are occurring, and determine the neutrino properties. In particular:
\begin{itemize}
\item It is very imporant to actually observe a neutrino oscillation (or at least a dip)
consistent with the characteristic $L/E$ dependence.
The dedicated MONOLITH experiment~\cite{antonioli}, or a future $\nu$ factory should be able
to do this cleanly. Long baseline experiments (K2K~\cite{hill}, 
MINOS~\cite{paolone}, ICARUS~\cite{rubbia,ereditato}) may also
observe oscillation patterns, as may the full (two detector) BOONE~\cite{spentzouris}.
\item One wants to determine the number of neutrinos and their nature, e.g., are there
sterile neutrinos, and are the active neutrinos Dirac or Majorana?
Mini-BOONE~\cite{spentzouris} should be able to confirm or eliminate the LSND results,
clarifying the need for sterile neutrinos.
$\beta \beta_{0\nu}$ may help resolve the Dirac/Majorana issue, although only for
some mass/mixing patterns.
\item The neutrino mass and mixing spectrum is important for its theoretical implications, for
choosing the solar neutrino solution, and for possible astrophysical implications. It will be
difficult to observe the  absolute scale of the spectrum, but some improvement may still be
possible from $\beta \beta_{0\nu}$ and from kinematic (laboratory~\cite{vissani} and,
especially, supernova~\cite{virtue}) observations. 
\item  It may be possible to confirm
that the atmospheric neutrino effects are mainly due to $\nu_\mu \RA \nu_\tau$ by
$\tau$ appearance in future long baseline experiments (or possibly in atmospheric
observations), but it will be harder to constrain small admixtures of $\nu_s$.
\item Leptonic CP violation is of theoretical interest. This may be observable (depending
on the parameters) at a neutrino factory or possibly a superbeam.
\end{itemize}

\section{COSMOLOGY~\cite{masiero}}

\subsection{Baryon asymmetry}
An important cosmological issue is the origin of the baryon asymmetry
\beq \Delta B \sim \frac{n_B - n_{\bar{B}}}{n_\gamma} \sim 10^{-9}. \eeq
In the standard model sphaleron solutions describe tunneling
between vacua with different $B$ at  temperatures $T$ above the electroweak
phase transition $T_{\rm EW} = $ O(100 GeV), which erase any preexisting
baryon asymmetry with $B - L =0$. This includes asymmetries
generated by the standard out-of-equilibrium decay scenario for heavy colored scalars
in grand unified theories. It is possible that an asymmetry is regenerated by the standard
model sphaleron effects at the time of the electroweak transition (e.g., in
processes
involving expanding bubble walls), but the necessary CP violation and out of equilibrium
constraints are not satisfied in the standard model or the MSSM.

One attractive scenario involves the initial generation of a nonzero lepton
asymmetry (or an assymetry in $B-L$) in the
early universe, which is then converted into comparable baryon and lepton asymmetries 
by sphalerons during the electroweak transition. For example, the heavy Majorana neutrino
$N$ expected in seesaw models may have asymmetric decays into a Higgs
scalar $h$ and
neutrino~\cite{leptogenesis}
\beq \Gamma (N \RA h \nu) \ne \Gamma(N \RA h \bar{\nu}) \eeq
if there is sufficient leptonic CP violation, leading to a nonzero $\Delta L$, which
is later converted to $\Delta B \ne 0$ and $\Delta L \ne 0$.
Such scenarios place constraints which are in principle stringent  
on the ($L$-violating) neutrino masses, to 
avoid wiping out both $\Delta B$ and $\Delta L$~\cite{leptogenesis},
typically $m_\nu < {\rm O}(1-10)$ eV, but with
large theoretical uncertainties. This applies to \nut \ and \num \ as well
as \nue.

\subsection{Relic neutrinos and mixed dark matter}
The active neutrinos decoupled just prior to big bang nucleosynthesis,
when the age of the universe was around 1s and the temperature around 1 MeV.
Their momentum distribution subsequently  redshifted to an  effective
temperature
$T_\nu \sim 1.9$ K, and they have an average density of around 300$/{\rm cm}^3$.
The direct detection of such low-energy neutrinos remains an ultimate challenge.

Massive neutrinos in the eV range will contribute a significant hot dark matter
(HDM) component to the
total matter of the universe. Constraints on the total energy density
imply
\beq \sum m_{\nu_i} \simle   35 \ {\rm eV}, \eeq
where the sum extends over the light, stable, active neutrinos, and also includes
light sterile neutrinos for some ranges of masses and mixings. However,
pure HDM models have long been excluded by observations of structure, since
neutrinos free-stream and produce large structures first, and there has not
been enough time for the observed smaller structures to form.
Until recently, mixed dark matter (MDM) models were very popular. In these,
neutrinos with \beq \sum m_{\nu_i} \simle  {\rm few \ \ eV} \eeq
contribute a hot component and account for large scales such as superclusters,
while a larger component of cold dark matter (CDM) explains structure on smaller scales.
These MDM models were a primary motivation for the degenerate 3 neutrino models
and an attractive aspect of 4 $\nu$ models. Most assumed 
$\Omega_{\rm matter} = \Omega_\nu + \Omega_{\rm CDM} = 1$, as expected in inflation models.

Recently things have changed dramatically, because:
\begin{itemize}
\item Most determinations now yield $\Omega_{\rm matter} \sim 0.3$.
\item Studies using Type IA supernovae as standard candles  suggest an accelerating universe,
with a cosmological constant (or other form of dark energy, such as quintessence),
with $\Omega_\Lambda \sim 0.7$.
\item Consistent and independent information comes from recent observations of
the location of the first Doppler peak in the cosmic microwave background radiation (CMB)
asymmetry spectrum, which implies $\Omega_{\rm matter} + \Omega_\Lambda \sim 1$.
\end{itemize}
There is therefore emerging a fairly compelling picture involving a low 
$\Omega_{\rm matter}$ and larger $\Omega_\Lambda$. This is consistent
with the expectations of inflation ($\Omega_{\rm matter} + \Omega_\Lambda = 1$), but the
evidence is purely observational. In this scenario, there is no need for a significant
component of HDM, although it is not excluded. 

Nevertheless, the observation of 
neutrino mass implies a small contribution to $\Omega_{\rm matter}$.
In particular, $m_3 \sim (\Delta m^2_{\rm atm})^{1/2}$ (hierarchical neutrinos) implies
$\Omega_\nu \sim 0.001-0.003$, while in the degenerate schemes
$\Omega_\nu$ could be as large as $\sim 0.1$.   Masses less than 
$m_\nu \sim 1 $ eV are not important for the observed structure, but may
be noticeable in the CMB spectrum for $m_\nu > 0.1$ eV.

\subsection{Big bang nucleosynthesis}
The  abundances of primordial $^4He$ and $D$ can be used to determine
the equivalent number $N_\nu^{\rm eff}$ of light neutrinos in equilibrium
at the time of neutrino decoupling ($t \sim $ 1 s, $T \sim 1$ MeV)
and the baryon density $h^2 \Omega_B$, where $h \sim 0.65 \pm 0.05$:
\beq 1.7 < N_\nu^{\rm eff} < 3.3; \ \ \ \  h^2 \Omega_B \sim 0.017(3). \label{nnubbn} \eeq
$N_\nu^{\rm eff}$ is actually an effective parameter, incorporating any
contribution to the $n/p$ ratio at the time of decoupling, i.e.,
\beq N_\nu^{\rm eff} = n_{\rm active}^{\rm eff}  + 
n_{\rm sterile}^{\rm eff} + n_{\rm asym}^{\rm eff},
\label{nnueff} \eeq 
where $n_{\rm active}^{\rm eff} \sim 3$ is from the active neutrinos, and $n_{\rm
sterile}^{\rm eff}$ represents the number of light sterile neutrino species present  at
decoupling. It has long been known that  sterile neutrinos
could be generated in equilibrium numbers by mixing with active neutrinos
prior to nucleosynthesis for a wide range of \dms \  and \stt.
In particular, it was believed that $n_{\rm sterile}^{\rm eff} \sim 1$
for mixing parameters corresponding to atmospheric $\nu_\mu \RA \nu_s$
oscillations, 
 in conflict
with (\ref{nnubbn}). (The recent Super K data also directly exclude pure
$\nu_\mu \RA \nu_s$). For the solar neutrinos, the rates have always allowed a
small mixing angle (SMA) $\nu_e \RA \nus$ solution analogous to the SMA $\nu_e \RA \num$,
but  in this case the mixing would {\em not} have been sufficient to generate \nus \
cosmologically\footnote{There is no viable large mixing angle (LMA) or low mass (LOW)
solar neutrino solution for $\nu_e \RA \nus$. 
These are most
likely excluded independently
 by the BBN data.}.

There has been considerable interest in ways to modify (or lower) the prediction 
for $N_\nu^{\rm eff}$, motivated by: (a) the observed value, which depends
on the somewhat controversial determinations of the relic $^4He$ abundance,
may eventually settle at a  value lower than 3. (b)
If there really is a light sterile neutrino, as suggested by the LANL data, 
then the possible contributions of $n_{\rm sterile}^{\rm eff}$ become
crucial. There are several canonical ways to change the prediction:
\begin{itemize}
\item 
If $\nu_\tau$ is unstable, then the active contribution to
$N_\nu^{\rm eff}$ may fall below 3, depending on the \nut \ lifetime and the 
energy density in its decay products. 
\item Large lepton asymmetries 
$\Delta L_a \simeq (n_{\nu_a} - n_{\bar{\nu}_a})/n_\gamma$
lead to increased energy densities $n_{\rm asym}^{\rm eff} > 0$, 
increasing  $N_\nu^{\rm eff}$. 
For $a=\mu, \tau$ this is the only effect. However, a positive
$\Delta L_e$ can actually decrease $N_\nu^{\rm eff}$ because it preferentially drives
the rate for $\nue n \RA e^- p$ compared to its inverse, and thus decreases the $n$ and
therefore the $^4He$ abundance. Neither of these effects are important
unless the asymmetry is very much larger than the baryon asymmetry $\sim 10^{-9}$.
\item
A massive $\tau$ neutrino in the range
between $\sim$ 0.5 MeV and the laboratory limit $\sim$ 18 MeV
would {\em increase} $n_{\rm active}^{\rm eff}$ above 3, because the large rest
energy would be more important than the reduced number density. This range
is therefore most likely excluded.
\end{itemize}

Foot and Volkas and others~\cite{fv} have recently reexamined the effects
of sterile neutrinos on nucleosynthesis, and argued that they could actually
{\em decrease} $N_\nu^{\rm eff}$. The current
status was discussed by Kirilova~\cite{kirilova} and Wong~\cite{wong}. There 
may be a strong
interplay between active-sterile mixing and a nonzero asymmetry $\Delta L_a$
(e.g., a small preexisting $\Delta L_a \sim 10^{-9}$ or one generated by the
oscillations). In particular, such effects can suppress the production
of \nus, deplete the number of \nue, distort the active neutrino spectrum
and therefore the reaction rates (a very important effect), and
amplify (or generate) a small initial $\Delta L_e$ asymmetry. The net result is that
the bounds on active-sterile mixing may be weakened, especially for small mixing.
However, there is still considerable debate as to the details and the size of 
these effects.

\subsection{Cosmic microwave background radiation}
Neutrino masses as small as 0.1 eV may lead to observable effects in the
CMB anisotropies. Pastor reviewed~\cite{pastor} the cosmological
implications of large lepton asymmetries, e.g., generated by an 
Affleck-Dine~\cite{affleck}
scenario or active-sterile mixing. In addition to the BBN effects, 
the asymmetry increases the radiation density and postpones the onset
of matter domination. This would suppress the CMB power
spectrum on small scales even for $m_\nu =0$, with larger effects for
$m_\nu \ne 0$. Masses as small as $10^{-2}$ eV might be observable in the
presence of an asymmetry.

Mangano~\cite{mangano} reviewed precision cosmology, emphasizing
the interplay between BBN, CMB, and large scale structure data.
He described the implications of
the recent MAXIMA and BOOMERANG CMB data, which indicate a suppressed second Doppler peak.
Several models of the cosmological parameters fit the new data, but
should be separable in future (MAP, PLANCK) data on the third peak.
One possibility is to increase the baryon density  $h^2 \Omega_B$ to a higher value
($0.030^{+0.012}_{-0.006}$) than is allowed by the BBN data
in (\ref{nnubbn}). One rather creative resolution would be to somehow
increase to $4 < n_{\rm active}^{\rm eff}  + 
n_{\rm sterile}^{\rm eff} < 13$, with 8 favored, and to simultaneously
assume a large $\Delta L_e \sim 0.24 \mu_{\nu_e}/T$, 
corresponding to $0.07 < \mu_{\nu_e}/T < 0.43$,
where $\mu_{\nu_e}$ is the \nue \ chemical potential.

\section{VIOLENT~ASTROPHYSICAL~EVENTS}
\subsubsection{High energy neutrinos}
There are many possible sources of ultra high energy astrophysical neutrinos,
including gamma ray bursts (GRB), active galactic nuclei (AGN), and 
particle physics exotica. They are a particularly useful probe because
high energy $\gamma$ rays tend to be absorbed in the astrophysical
source, while protons are magnetically deflected. Also, protons
above $10^{20}$ eV can be
observed only from local sources because of the Greisen-Zatsepin-Kuzmin (GZK) cutoff
(scattering from the CMB).

Waxman~\cite{waxman}, Halzen~\cite{halzen}, and Perrone~\cite{perrone} described the
theoretical expectations for GRBs, and the need for km$^3$ scale detectors.
GRBs are now understood to be mainly at  cosmological distances,
and arise from the expansion of a relativistic fireball.
The expanding shock accelerates protons and  high energy $e^-$, which emit
the $\gamma$'s by synchrotron emission. Outstanding issues are the fireball progenitor
(e.g., coalescence of neutron stars or of a neutron star and black hole)
and the $e-p$ coupling mechanism.

Reactions such as $\gamma p \RA \pi^+ n$ from the GRB photons are expected to produce
a burst of 100 TeV $\nu$'s, followed by the production of $10^{18}$ eV $\nu$'s for about 10 
s associated with the expansion of the shock into the interstellar medium (the afterglow).
$p n \RA \pi^+$ processes also lead to lower energy (10 GeV) neutrinos.
Estimates of the fluxes suggest the need for a km$^3$ detector
for a reasonable event rate of order 10's/yr. One also expects a few/yr
from the diffuse scattering of protons from the CMB photons.
The $\gamma$'s from $\pi^0$ decay are
absorbed and cascade to lower  energies $\simle$ TeV. These may be ultimately observable.

The observed ultra high energy cosmic ray protons may also be due to this mechanism
(those above  $10^{20}$ would have to be due to nearby GRBs to evade the
GZK cutoff). One can turn this argument around: any mechanism that produces
high energy $\nu$'s from $\gamma p$ (or $p N$) interactions in an
optically thin source should produce high energy $p$'s as well as $\nu$'s.
The observed $p$ flux therefore limits the possible $\nu$ flux (the Bahcall-Waxman
bound), invalidating some early optimistic estimates. The bound can be evaded in
optically thick sources, but most plausible models are thin.

There are other possible sources of high energy $\nu$'s. These include
the annihilation of WIMPs in the earth, the sun, or the galactic center;
monopole annihilation; the decay of topological defects 
(lattice calculations are being carried out by the TRENTO group~\cite{gibilisco});
or local astrophysical sources. High energy neutrinos may also produce
$Z$'s by annihilation with the relic neutrinos.

One spectacular prediction~\cite{waxman}
 for oscillations is that an initial \num \ source
should yield equal numbers of \num \ and \nut \ for $\dms > 10^{-17}$ eV$^2$ for
maximal mixing ($\stt \sim 1$).

The present experimental status and future prospects were reviewed
by Halzen~\cite{halzen}, Vignaud~\cite{vignaud},
van Dantzig~\cite{vandantzig}, and de Marzo~\cite{demarzo}.
The MACRO and Lake Baikal detectors had areas $A < 10^3$ m$^2$.
(Baikal observed some events). In the $A \sim 10^4$ m$^4$ range,
AMANDA (S. Pole) has been running for some time, and observed 193 atmospheric
$\nu$'s (it has mainly vertical sensitivity), and NESTOR (near Greece)
is under development. The larger AMANDA II with $A \sim 10^5$ m$^2$
is taking data. It may eventually be succeeded by the km$^3$ ICE-CUBE.
Also in the $A \sim 10^5$ m$^2$ range is ANTARES~\cite{vandantzig}
 in the Mediterranean,
which should run in 2002-2003. This should be succeeded by ANTARES II.
Another project, NEMO~\cite{demarzo} is under study. It could be deployed
near Sicily, or merge with ANTARES II.

\subsubsection{Supernova implications}
A Type II supernova is due to the collapse of the iron core of a star
with  mass exceeding  $\sim 8 M_\odot$.
The core collapses into a neutron star or black hole. The initial collapse
leads to a ms neutronization pulse of \nue \ from $ e^- p \RA \nue n$.
The collapsing core eventually bounces, with an expanding shock, leaving
behind a dense hot core and neutrinosphere. The latter radiates neutrinos
of all types over a period of $\sim$ 10 s. The characteristic
temperature of the $\nu_\mu, \bar{\nu}_\mu,  \nu_\tau, \bar{\nu}_\tau$
is $\sim $ 8 MeV. The \nue \ and $\bar{\nu}_e$ stay in equilibrium longer
due to charged current interactions with matter, implying smaller temperatures,
e.g., $T_{\nu_e} \sim 3.5$ MeV, $T_{\bar{\nu}_e} \sim 4.5$ MeV~\cite{haxton}.
Neutrinos are relevant because:
\begin{itemize}
\item Almost all (99\%) of the
energy ($\simgr 3 \x 10^{53}$ ergs) is radiated in neutrinos.
The spectacular optical effects are a perturbation.
\item Observation of a neutrino burst may give an early warning
of a supernova, with the Solar $\nu$ Early Warning System (SNEWS)
 network under organization~\cite{virtue}.
\item Neutrinos are important for the dynamics. Scattering of neutrinos radiated from
the neutrinospere may revive a stalled shock, leading to the observed explosion.
\item The $\nu$-heated supernova ejecta is a favored  candidate for the
site of the $r$-process, which refers to the synthesis of nuclei heavier
than iron by the rapid capture of neutrons on a heavy core in a neutron-rich environment.
However, some estimates~\cite{fuller} suggest that 
 $  \nue n \RA  e^- p$, with the $p$ immediately incorporated into
an $\alpha$, will be too efficient at
destroying neutrons, preventing the $r$-process. The situation can be worsened or improved in
the presence of neutrino mixing.
\item The kinematic effects of neutrino mass can distort the time and energy spectrum
of the neutrinos. The observed Kamiokande and IMB events from SN 1987A,
which were sensitive to $\bar{\nu}_e$ from the neutrinosphere,
allowed a limit of around $m_{\nu_e} \simle 20$ eV. This limit, as well as many
other constraints, depended on the theoretical modelling of the supernova.
 \item A future supernova within our galaxy should yield large numbers of events
in large detectors if they are running. This should allow
much more stringent direct limits on $m_{\nu_\mu, \nu_\tau}$
than by any laboratory method~\cite{virtue}. For example, SNO should be sensitive to
30 eV, and Super K to 50 eV neutrinos~\cite{beacom}. 
A collapse into a black hole would provide a sharp cutoff in time for the
neutrino signal, allowing even more
precise constraints (in  the few eV range) for all neutrino types~\cite{bh}.
With large numbers of
events it will be possible to study the supernova dynamics in detail and
to constrain other neutrino properties. SNO should be especially useful because
it can separately observe \nue \ , $\bar{\nu}_e$, and the neutral current
scattering of all neutrinos.
\item Neutrino mixing can lead to a variety of oscillation and MSW resonance effects.
Because the densities are higher than in the Sun, there may be resonant conversions for
higher \dms \ than the solar neutrinos. In particular,
  \begin{itemize}
   \item $\nu_e \leftrightarrow \nu_{\mu, \tau} $ conversions can increase the
final \nue \ energy because of the harder initial $\nu_{\mu, \tau} $ spectrum.
This makes \nue \ scattering more efficient in reviving the stalled shock.
On the other hand, it aggravates the problem of destroying neutrons before
they can participate in the $r$-process, excluding $\dms > $ few eV$^2$ escept
for very small mixing~\cite{fuller}.
  \item For the LMA solar neutrino solution, there may have been a
partial conversion of $\bar{\nu}_e$ and $\bar{\nu}_\mu$,
in contrast with the observed SN 1987A spectrum~\cite{bss,cline}.
However, it has recently been argued that matter effects may reduce
this difficulty, or even help reconcile the observed Kamiokande and
IMB spectra~\cite{lunardini}.
\item Minakata~\cite{minakata} argued that the observed spectra would
be very different for an inverted spectrum, e.g., with the 
the dominant \nue \ mass eigenstate heavier than \nut.
This could lead to a determination of the sign of 
$\Delta m^2_{23}$.
\item It has been argued that active-sterile conversions could
solve the $r$-process problem~\cite{caldwell} by the sequence of
$\num \RA \nus$ followed by $\nue \RA \num$.
  \end{itemize}
\item One expects a supernova in our galaxy on average every 30-100
years. This is an unfortunate mismatch with the practical human time scale
for carrying out experiments, but we should make an attempt.
Large neutrino detectors should be designed to run for a minimum
of 10-20 yr, and preferably longer.
\end{itemize}

\section{LOW ENERGY NEUTRINOS}

\subsubsection{Solar neutrinos~\cite{krastev}}
The Solar neutrinos gave the first convincing evidence for neutrino
mass and mixing.
Bahcall reviewed~\cite{bahcall} the status of the standard solar model (SSM),
emphasizing that ``Solar model predictions have been robust for 30 years.''
\begin{itemize}
\item The new Bahcall-Pinnsoneault (BP00) code for the standard solar model
incorporates improved
opacities (at the edge of the table); minor nuclear refinements; 
a detailed electron density profile $n_e(r)$; a denser model grid,
improved treatment of helioseismology constraints; and detailed studies of
the time dependences in the solar radius, luminosity, and radius of the
convective zone. These latter will be measured for solar-type stars
(e.g., by interferometry), testing a new aspect of the solar models.
\item Helioseismology now confirms most aspects of the standard 
solar model (except some nuclear cross sections and the neutrino properties)
in more than sufficient detail for the interpretation of solar neutrino
data~\cite{bahcall,villante}.
\item The data indicate in a model independent way that the
suppression of $^7Be$ neutrinos is much greater than
the $^8B$ suppression. This implies, independent of any specific solar
model, that the explanation of the solar neutrino anomaly cannot
be due to astrophysics. (It is still very useful to use the standard solar
model results as the starting point for the parameters and uncertainties
in MSW and vacuum oscillation analyses.)
\item Changes in the predictions from BP98 
include a +4\% increase in the $^8B$ flux, and
a 2\% increase for $^7Be$. This leads to an increase of 1 SNU (to $130^{+9}_{-7}$)
for $^{71}Ga$, and of 0.3 SNU (to $8.0^{+1.2}_{-1.0}$) for $^{37}Cl$.
\item The largest residual theory uncertainties are from nuclear cross sections.
  \begin{itemize}
   \item The largest uncertainly is still from the
$^7Be(p,\gamma)^8B$ reaction. Four new experiments are expected to reduce
the uncertainty to the 5-10\% range. One would  prefer 5\%.
   \item The next most important uncertainty is from $^3He(\alpha,\gamma)^7Be$,
for which there are experimental discrepancies.
   \item There is still no firm uncertainty in the predicted hep ($^3He +  p \RA
^4He +  e^+ \nu_e$) flux. With improved energy resolution in the SuperK
experiment this can be determined independently from events beyond the $^8B$ endpoint of 
$\simgr$ 14 MeV, so it is no longer so crucial in the interpretation of the 
energy spectrum.
   \item Kubodera~\cite{kubodera} reviewed the theoretical status of the
$\nu_e d \RA e^- pp$ and $\nu_X d \RA \nu_X p n$ ($X = e, \mu, \tau$)
cross sections, which will be needed for the interpretation of the SNO results.
He described a new potential model calculation in which the potential and
exchange currents were calculated consistently, and also a one-parameter
effective field theory treatment which is in good agreement.
The conclusion is that the theory errors are under control, especially for
the (neutral current)/(charged current) ratio.
  \end{itemize}
\end{itemize}

Ferrari~\cite{ferrari} described the results of Run I of the
Gallium Neutrino Observatory (GNO). They obtain a
rate of $65.8^{+10.2+3.4}_{-\ 9.6-3.6}$ SNU, consistent with
the earlier GALLEX and SAGE results, and about half of the SSM
prediction.

Suzuki~\cite{suzuki} presented new solar neutrino results for Superkamiokande:
\begin{itemize}
\item The current results are based on 1117 d of running, compared to
the 824 d reported last year. There is a unified new analysis of all
of the data.
\item The rate (compared to the BP98 prediction) is $0.465 \pm 0.005^{+0.015}_{-0.013}$.
\item The energy threshold has been lowered to 5 MeV. However, the 5-5.5
MeV bin is not yet used in the oscillation analysis.
\item The recoil spectrum (compared to the SSM) is flatter than before,
and is now consistent with flat ($\chi^2/df = 13.7/17$). 
The first moment of the recoil energy  is $8.14 \pm 0.02$ MeV.
Distortions are mainly expected for the SMA and vacuum solutions.
\item The small excess of events at night (N) compared to day (D) has been reduced from 
$1.8\sigma$ to $1.3\sigma$, i.e.,
\beq  \frac{N-D}{(N+D)/2} = 0.034 \pm 0.022^{+0.013}_{-0.012}.  \eeq
Day-night differences may be important for the LMA and LOW solutions.
\item No seasonal variation (as expected for vacuum oscillations) has been
observed beyond the $1/r^2$ effect. The seasonal data has not yet been fit
quantitatively.
\item A fit in which the hep flux is left free yields a flux $5.4 \pm 4.5$ 
compared with BP98. Leaving the flux free has no significant effect on the
oscillation analysis.
\item There is no evidence for long term time variation (as expected in
RSPF models with transitions in the convective zone).
\end{itemize}

Suzuki presented the SuperK oscillation analysis (including the rates from
other experiments). He argued that there is no smoking gun for one solution 
over another. However, the data favors the LMA solution; the SMA solution
is disfavored but not excluded; vacuum solutions are disfavored;
and the pure $\nu_e \RA \nu_s$ is disfavored at 95\%,
by a combination of day-night, spectrum, and rate information. (A significant
admixture of \nus \ is still allowed in 3$\nu$ fits~\cite{garcia}, 
especially for small mixing angles.)

Smirnov~\cite{smirnov} discussed the various solar neutrino solutions,
commenting that ``Nature selects the most ambiguous solution'', i.e.,
no solution is strongly excluded or preferred, and hints are at the level of
systematics. He argued that the LMA is favored, but SMA may be back.
Smirnov described that the SuperK zenith spectrum has a small excess in the
first night-time bin, contrary to any of the solutions (LMA is flat at night,
LOW may yield an excess in the second bin, and SMA may imply an
excess in the last (core) bin), adding to the confusion. He
argued the importance of studying the correlations between various observables
(e.g., charged current rate vs. day-night) for distinguishing 
solutions in the future.

A number of global analyses of the solar data were presented at the conference or
in the recent literature~\cite{lisi,smirnov,suzuki,garcia}, allowing mixing between
2, 3, or 4 neutrinos, including sterile. 
The 2$\nu$ solutions are shown schematically in Figure~\ref{bigpicture}.
The small mixing angle (SMA) solution, with
$\dms \simle 10^{-5}$ eV$^2$ and $10^{-3} < \stt < 10^{-2}$,
 is most analogous to the (small)
quark mixings. The large mixing angle (LMA) and low mass (LOW) solutions,
with $\dms \sim (10^{-5}-10^{-4})$ and $10^{-7}$ eV$^2$, respectively,
are close to the vacuum solutions ($\stt = 1$). Regeneration in the earth
is important in these solutions. The original vacuum (VAC) solutions
with $\dms \sim 10^{-10}$ eV$^2$ involved 
an accidental coincidence between the Earth-Sun distance and the vacuum 
oscillation length,
and therefore predicted large energy and seasonal variations that are
now  mainly excluded.
Matter effects are starting to be relevant in the transition or  
 quasi-vacuum region~\cite{montanino}, ($10^{-10} - 10^{-7}$ eV$^2$).
There is a SMA solution for transitions  to sterile neutrinos, but
no viable analogs of the other regions. The major difference compared with
active neutrinos is the absence of a neutral current component
to $\nu_e e \RA \nue e$ in Kamiokande and Superkamiokande.
There is a smaller effect from the small neutron density in the Sun.

All of the analyses, which had
slightly different inputs, favor the LMA solution,
with the SMA, LOW, vacuum, and pure sterile (with parameters in the
SMA range) solutions
disfavored  to various extents, depending on the analysis. One of the differences was
in the simultaneous application of the SuperK spectrum and zenith angle
distributions. The full two-dimensional distribution with correlations has never 
been presented, so each analysis has either used separate distributions
(which double counts the same data), or ignored some of the data.
Another problem or source of ambiguity is the statistical treatment of
non-Gaussian effects, such as the
allowed regions of parameters when there are multiple
minima.

Many of the recent studies have used $\tan^2 \theta$ rather than \stt \
as the mixing parameter. Allowing  $\tan^2 \theta > 1$ includes the
second octant $\pi/4 < \theta \le \pi/2$. This ``dark side''~\cite{dark} was
usually ignored in  past studies, which assumed implicitly that
$0 \le \theta \le \pi/4$ and \dms $>$ 0. (The second octant
solutions correspond to exchanging \nue \ and the second neutrino, i.e., 
they are equivalent to restricting $\theta \le \pi/4$ but allowing \dms $< 0$.)
The effect is mainly important for the LOW solutions, 
which extend well into the dark side~\cite{garcia}. However, LMA
does also at high confidence level. Vacuum solutions do not depend 
on the sign of \dms, so
they are mirror-symmetric.

There are many future experiments underway or proposed:
\begin{itemize}
\item Noble ~\cite{noble} described the status of  SNO.
The Phase I, charged current (CC), running is near complete, and
the collaboration may publish the CC flux soon. When combined with
the SuperK electron scattering (ES) results, this would allow
an indirect determination of the (neutral current (NC))/CC ratio,
and therefore of the \nus \ flux. It remains to be determined whether
the systematics would allow a meaningful indirect determination. 
In any case, SNO will then proceed to Phase II, in which salt is added to allow
a  clean direct measure of the NC/CC ratio.
SNO will also measure the CC recoil spectrum. Although the statistics will be lower than
SuperK, SNO has the  advantage that the $e^-$ energy is a direct measure of the \nue 
\ energy, i.e., there is no convolution. They will also be sensitive
to day/night and seasonal effects, and be an excellent supernova detector.
\item The Borexino experiment~\cite{meroni} will determine
the $^7Be$ flux (actually, a combination of the \nue \ and converted $\nu_{\mu,\tau}$
flux), as well as day/night and seasonal effects.
\item Proposed future experiments~\cite{dekerret,laubenstein}, including
HELLAZ, HERON, LENS,  MOON, and GENIUS, would be sensitive to the $pp$ and $^7Be$
neutrinos. Their day/night and seasonal sensitivities would be especially useful
for the LOW solution.
\item  KAMLAND~\cite{gratta} is a long baseline reactor $\bar{\nu}_e$
disappearance experiment in the Kamiokande mine. It would be sensitive
to most of the power reactors in Japan, and should probe into the LMA range.
It may also be used to observe solar $^7Be$ neutrinos.
\item The proposed GENIUS~\cite{klapdor}  $\beta \beta_{0\nu}$
experiment could possibly be sensitive to masses in the (hierarchical) LMA range,\
but only with a sensitivity to $\langle m_{\nu_e} \rangle \sim 10^{-3}$ eV
(the anticipated sensitivity is $2 \x 10^{-3}$ for a 10 ton detector run for 10 yr).
\end{itemize}

The theoretical and experimental situation for the solar neutrinos has
become very mature. Refined and multiple experiments are needed
to simultaneously determine the neutrino solutions (especially if nature
chooses a complicated scenario, e.g., with 3 relevant neutrinos) and constrain the
astrophysics. As we enter into a ``precision'' phase with several 
high statistics experiments
with multiple observables, it is important that the data be presented and 
analyzed in a manner that allows us to extract the maximum and most reliable conclusions.
In particular, it is important for each group to publish all of their data
with fully correlated errors (e.g., the spectral and zenith data should
be presented as a two-dimensional distribution). Similarly,
to ensure an optimal treatment of common sytematics and theory uncertainties,
it would be useful for the experiments to perform combined analyses,
analogous to the highly successful LEP Electroweak Working Group.

\subsubsection{Atmospheric neutrinos~\cite{lipari,learned}}
The theoretical flux calculations were reviewed by
Stanev~\cite{stanev}, Battistoni~\cite{battistoni}, and Lipari~\cite{lipari}.
The largest uncertainties are in the primary flux and in the pion yield
in $p$-nucleon scattering. The agreement between the Bartol and
Honda flux calculations was somewhat accidental, since the Bartol group used
a lower flux and higher $\pi$ yield, while Honda assumed the reverse.
Recent new measurements of the primary flux by AMS~\cite{bertucci} 
and BESS~\cite{sanuki} have 
clarified the situation up to a few hundred GeV (the range relevant for fully
contained and for partially contained or stopping events), indicating that
the old (high) ``Weber data'' should be discarded. There is still a need
for new measurements at middle energy (up to a few TeV), relevant
to throughgoing muons. There has been some theoretical work on the
pion yields, but major progress is expected from the HARP cross section
measurements at CERN~\cite{catanesi} (which are also critical to a possible $\nu$ factory).

In addition to the Bartol and Honda calculations, the newer FLUKA group~\cite{battistoni}
has explored theoretically-motivated (as opposed to semi-empirical) cross sections.
Moreover, they have done a new 3-dimensional calculation. This yields somewhat
higher fluxes in the horizontal ($\cos \theta \sim 0$) direction than the older
one-dimensional codes, leading to slightly higher \dms. 
Many smaller effects, including geometry, geomagnetic effects, seasonal variations,
etc., have been carefully considered.

The conclusion is that the flux is well understood. There are no large problems, and
the predicted $\nu_\mu/\nu_e$ ratio and zenith distributions are under control.
All of the groups are
working on refined calculations.

Ronga~\cite{ronga} reviewed the data from Soudan 2 and MACRO.
Soudan 2 obtains a $\nu_\mu/\nu_e$ ratio of 0.68(11)(6) compared to
expectations, and their up/down asymmetry is becoming significant.
MACRO excludes  pure $\nu_\mu \RA \nus$ at 98\%, mainly from the ratio
of vertical to horizontal upward throughgoing events.
(There are matter effects for $\num \RA \nus$, or for
$\num \RA \nue$, but not for $\num \RA \nut$.)

Kajita~\cite{kajita} summarized the current status of Superkamiokande data:
\begin{itemize}
\item The zenith angle distributions are now very precise and beautiful.
They are consistent with oscillations. However, they do not have the sensitivity
to resolve oscillation wiggles, so neutrino decay scenarios cannot be excluded.
\item Analysis of the data using the new 3d flux calculations will increase
\dms \ slightly from the value ($\sim 3.2 \x 10^{-3}$ eV$^2$) obtained
using a 1d flux. The precise number is under investigation. The higher \dms \
is good news for long baseline experiments.
\item Pure $\nu_\mu \RA \nus$ is excluded at 99\% by a combination
of upward throughgoing, high energy PC, and NC-enhanced multi-ring events.
However, significant admixtures of \nus,  such as $\nu_\tau + \nus$,
in the final state cannot be excluded.
\item SuperK has fit to a transition probability $\propto \sin^2 (\alpha LE^n)$, with
$\alpha$ and $n$ free. They obtain $n = -1.0 \pm 0.14$,
consistent with oscillations ($n=-1$), but excluding CPT violation ($n=0$),
or violations of Lorentz invariance or the equivalence principle ($n=+1$).
\end{itemize}

In the future there will be refined flux calculations. It is 
possible that SuperK will be able to observe $\tau$ appearance at the 2-3$\sigma$ level,
further
constraining the sterile neutrino scenarios. Atmospheric neutrinos
may be studied in new generation experiments, including 
MONOLITH~\cite{antonioli} and ICARUS~\cite{rubbia}  at
Gran Sasso and the ANTARES underwater experiment~\cite{carloganu}.
MONOLITH should have excellent capabilities for observing an oscillation dip,
as well as matter effects~\cite{chizov}.
Finally, terrestrial long baseline experiments K2K (KEK to Kamiokande)~\cite{hill},
MINOS (Fermilab-Soudan)~\cite{paolone}, and CNGS (Cern-Gran
Sasso)~\cite{rubbia,ereditato,bettini} will be able to study the atmospheric neutrino
parameter range in much more detail.

\subsubsection{Laboratory oscillation experiments~\cite{saitta}}
\begin{itemize}
\item The CERN short baseline (SBL)  experiments NOMAD~\cite{sitjes}
and CHORUS~\cite{zucchelli} are appearance experiments searching
for $\nu_\mu \RA \nut$ and, with less sensitivity, to $\nue \RA \nut$,
in the region of large \dms \ and small \stt \ that was especially
motivated by mixed dark matter scenarios. The NOMAD limits are almost
final, and CHORUS has completed Phase I. They imply $\stt < (4-6) \x 10^{-4}$
for large \dms.
\item LSND~\cite{spentzouris} is the only oscillation experiment with a
positive appearance signal. Their final transition probability, based on
both $\num \RA \nue$ and $\bar{\nu}_\mu \RA \bar{\nu}_e$,
is $P(\num \RA \nue) = (0.25 \pm 0.06 \pm 0.04)$\%. This is slightly lower
than their earlier results, reopening the possibility of $3+1$ scenarios
(Section \ref{scenarios}).  KARMEN 2~\cite{steidl} sees no
evidence for oscillations. However, there is a small region of parameters
($\dms \sim (0.2 - 1)$ eV$^2$, $\stt \sim(10^{-3}-10^{-2}$) for which
they are consistent when analyzed the same way. The LSND, atmospheric,
and solar neutrino results imply three distinct values of \dms, most
likely requiring
four neutrinos. Since the $Z$ lineshape only allows three light active neutrinos, this
would imply the need for a light sterile neutrino mixing with the active ones.
It is therefore crucial to confirm (or not) the LSND results. The miniBoone
(or later two-detector version, Boone) at Fermilab will be sensitive to the
LSND range~\cite{spentzouris}. An additional experiment would be useful.
\item The reactor long baseline experiments~\cite{gratta} CHOOZ and
Palo Verde exclude $\bar{\nu}_e$ disappearance for $\stt \simgr 0.1$ down
to $\dms \sim 10^{-3}$ eV$^2$. They therefore excluded $\num \RA \nue$ as the
dominant effect in atmospheric neutrinos. (The
\nue \ zenith distributions from SuperK excluded this possibility independently.)
However, \nue \ could still play a subleading role.
The future Kamland experiment in Kamiokande~\cite{gratta}
will have a sensitivity to reactors within several hundred  km.
It should be sensitive to \stt \ down to $(1.3-3)\x 10^{-1}$ for $\dms \ge 10^{-5}$
eV$^2$, i.e., all of the solar LMA range. There is also a Krasnoyarsk
proposal to go down to $\stt \sim$ few $\x 10^{-2}$.
\item Accelerator long baseline experiments are designed to search for \num \
disappearance and/or \nut \ or \nue \ appearance in the atmospheric neutrino range
 in a cleaner laboratory environment.
The K2K (KEK to Kamiokande) experiment~\cite{hill}, which is already running, looks
for \num \ disappearance. It is sensitive to the upper part of the atmospheric \dms \ range.
MINOS-NUMI (Fermilab-Soudan)~\cite{paolone} will search
for disappearance in the first phase, while the CNGS (Cern-Gran Sasso)
program (OPERA and ICARUS)~\cite{rubbia,ereditato,bettini} 
will concentrate on \nut \
appearance. They should cover most of the atmospheric range.
\end{itemize}

\subsubsection{More distant possibilities}
Blondel~\cite{blondel} and Gavela~\cite{gavela} described the
possibilities for a future neutrino factory, which has been discussed for
CERN or Fermilab~\cite{autin,spentzouris}. A neutrino factory refers to  intense and precisely
understood
\nue, \num, $\bar{\nu}_e$, and $\bar{\nu}_\mu$ beams produced at
a dedicated muon storage ring. Compared to more convention alternatives,
a $\nu$ factory would produce the best physics, but would be very expensive
and require a long time scale. However, it could be the first step towards a muon collider.

Dydak~\cite{dydak} compared and contrasted the possibilities of a $\nu$ factory
with those of a superbeam, which refers to a much more intense (e.g., by 100)
version of the
conventional beams from $\pi$ and $K$ decay. A superbeam could be built, for example,
at the Japan Hadron Facility (JHF), or at CERN or Fermilab. A superbeam  might
be much less expensive and faster to build than a $\nu$ factory, and might be
built first or instead. However, given the nature of conventional beams
(less well understood energies, \nue \ contamination)
the detector needs might increase the costs unacceptably.
In particular, the 1\% \nue \ contained in a conventional wide band
beam might be fatal for oscillation studies. It has been suggested that
this could be reduced by using a low energy narrow band beam 
generated by low energy (1-2 GeV) protons~\cite{richter}, but the rate
is reduced by 100, and there is still a 0.1\% \nue \ contamination from $\mu$ decay.
There are also uncertainties in the flux estimates. Another alternative is to
use low energy (few hundred MeV) \nue \  produced by the PRISM muon 
source~\cite{sato}, which
might allow the observation of CP violation without needing matter effects.

The motivations for a neutrino factory (or conventional alternative) are especially strong
if the LMA solution for the solar neutrinos turns out to be correct.
The goals include:
\begin{itemize}
\item A precise determination of the atmospheric parameters $U_{\mu 3}$ and
$|\Delta m^2_{23}|$.
\item A measure of the admixture $U_{e3}$ of \nue \ into $\nu_3$,
with a sensitivity down to $|U_{e3}|^2 \sim 10^{-3}-10^{-4}$.
\item The sign of $\Delta m^2_{23}$, which distinguishes the hierarchical and
inverted models, can be determined provided $|U_{e3}|^2 \ne 0 $ can be observed.
\item The  CP-violating phase $\delta$ in the leptonic mixing might be 
observable, provided that the parameters correspond to the LMA with
$\Delta m^2_{12} > 2 \x 10^{-5}$ eV$^2$ and $|U_{e3}|^2 \ne 0 $.
\item It will be necessary to determine $\delta$, $|U_{e3}|^2$, and matter
effects simultaneously. They can in principle be separated by exploiting
their different $L/E$ dependences. In principle one could use the $E$ dependence
at a single detector, but in practice one needs 2 or preferably 3 detectors
to handle systematics. Typical distances considered include $\sim 700$ km 
(Cern-Gran Sasso), 3000 km (Fermilab to California or Cuba, Cern to Canary Islands
or northern Norway), or 7300 km (Fermilab-Gran Sasso). The latter would require an
extremely challenging  nearly
vertical beam.
\item The physics possibilities would be much easier and richer in most 4 $\nu$ schemes.
\item One could carry out a fine program of other physics, such as deep inelastic
scattering.
\end{itemize}

Learned~\cite{learned} emphasized that future large-scale oscillation experiments
(e.g., the far detector at a neutrino factory) should also be designed to search
for proton decay, with the goal of  sensitivity to a $10^{35}$ yr lifetime.

\subsubsection{Non-oscillation experiments}

Direct kinematic limits on neutrino mass were reviewed by Vissani~\cite{vissani}.
At present $m_{\nu_e} < 2.2$  eV from Mainz and Troisk, possibly 
improvable to around 0.4 eV. There is little room for laboratory
improvement on $m_{\nu_\mu}$ or $m_{\nu_\tau}$. We must wait for
a supernova.

Klapdor~\cite{klapdor} reviewed neutrinoless double beta decay 
($\beta \beta_{0\nu}$), which can be driven by Majorana neutrino
masses, as well as by other mechanisms involving supersymmetry with $R$-parity
violation, heavy Majorana neutrinos,
leptoquarks, extended gauge interactions, compositeness, or the
violation of Lorentz invariance or the equivalence principle.

For light Majorana neutrinos, the amplitude is proportional
to $\langle m_{\nu_e} \rangle$, defined in (\ref{mnueff}).
The current upper limit of 0.35 eV at 90\% is from the Heidelberg-Moscow
experiment. Some authors would allow larger values because of theoretical uncertainties
in the nuclear matrix elements. There can be cancellations between the
contributions of the individual terms in (\ref{mnueff}), so that
$\langle m_{\nu_e} \rangle$ can be smaller than the mass eigenvalues.
In fact, a Dirac neutrino, for which $\langle m_{\nu_e} \rangle = 0$,
can be usefully be thought of as two degenerate Majorana neutrinos with
equal $U_{ei}$ and opposite CP-parities.

The existing limit eliminates the degenerate 3$\nu$ scenarios motivated by
mixed dark matter, unless they are Dirac or there are finely tuned cancellations,
and it excludes some of the 4$\nu$ schemes.
There are proposals to extend the $\langle m_{\nu_e} \rangle$
sensitivity considerably. These
include MOON (0.03 eV), EXO (0.01-0.03 eV)~\cite{gratta}, CUORE (0.05 eV),
and GENIUS~\cite{klapdor}.  The latter is a proposed enriched $Ge$ detector
shielded by liquid nitrogen. A one ton version could reach 0.02 (0.006)
eV in 1 (10) years, with 0.02 a typical expectation for the inverted model. 
A later 10 ton
version could reach $2 \x 10^{-3}$, slightly above the 0.001 needed to probe
the LMA hierarchical solution. GENIUS could also search for WIMPs
and could detect solar $pp$ and $^7Be$ $\nu$'s in real time.

\section{OUTLOOK}
With the direct observation of the \nut \ by the DONUT experiment~\cite{okada}
the standard model is complete except for the Higgs boson. Neutrinos
played a significant role in establishing the standard model, while
the observation of neutrino mass has also provided the first evidence
for new physics. The evidence for neutrino mass and mixing is now
convincing. New generations of experiments should establish the
neutrino mass and mixings, hopefully determine their nature (Dirac or
Majorana) and number, and perhaps establish leptonic CP violation.
Not only will this be a superb constraint on what underlies the standard
model, but neutrinos are also a critical probe of astrophysical
processes, from stars to gamma ray bursts and the universe as a whole.

\section{ACKNOWLEDGMENTS}
It is a pleasure to thank the conference organizers, particularly
Gianluigi Fogli and Eligio Lisi, for organizing such a successful meeting,
as well as the speakers who did such an excellent job summarizing this
exciting field. This work was supported by the U.S. Department
of Energy grant DOE-EY-76-02-3071 and by the workshop organizers.

\end{document}